\documentstyle[equations,12pt]{article}
\begin{document}

\begin{flushright}
IPNO/TH 96-21
\end{flushright}
\vspace{1 cm}
\begin{center}
{\large {\bf Incorporation of anomalous magnetic moments in the two-body
relativistic wave equations of constraint theory}}
\vspace{1.5 cm}

H. Jallouli\footnote{e-mail: jallouli@ipncls.in2p3.fr}
and H. Sazdjian\footnote{e-mail: sazdjian@ipno.in2p3.fr}\\
\renewcommand{\thefootnote}{\fnsymbol{footnote}}
{\it Division de Physique Th\'eorique\footnote{Unit\'e de Recherche des
Universit\'es Paris 11 et Paris 6 associ\'ee au CNRS.},
Institut de Physique Nucl\'eaire,\\
Universit\'e Paris XI,\\
F-91406 Orsay Cedex, France}\\
\end{center}
\newpage

\begin{center}
{\large Abstract}
\end{center}
Using a Dirac-matrix substitution rule, applied to the electric charge, 
the anomalous magnetic moments of fermions are incorporated in local form 
in the two-body relativistic wave equations of constraint theory. The 
structure of the resulting potential is entirely determined, up to
magnetic type form factors, from that of the initial potential descibing
the mutual interaction in the absence of anomalous magnetic moments.
The wave equations are reduced to a single eigenvalue equation in the
sectors of pseudoscalar and scalar states ($j=0$). The requirement of
a smooth introduction of the anomalous magnetic moments imposes restrictions
on the behavior of the form factors near the origin, in $x$-space. The 
nonrelativistic limit of the eigenvalue equation is also studied.
\par

\newpage

\section{Introduction}

The two-body relativistic wave equations of constraint theory have the
main feature of describing the internal dynamics of the system by means
of a manifestly covariant three-dimensional formalism 
\cite{ll,dv,t,t1,k,ls,cva1,s1,ms}; relative energy and relative time variables 
are eliminated there through constraint equations. These wave equations,
which can be constructed from general principles, have also the property 
of allowing a three-dimensional reduction of the Bethe-Salpeter equation
by means of a Lippmann-Schwinger-quasipotential type equation 
\cite{ltlttk,bs,g,pl,f,fh,t2,lcl,mw} that relates the two-body potential
to the scattering amplitude \cite{s2,bb}. In this way, the potential
becomes calculable, in perturbation theory, from Feynman diagrams.
\par
In a recent work \cite{js}, we applied this calculational method to the
evaluation, in certain approximations, of the potentials in the cases of
scalar and vector interactions, mediated by massless photons. It turns
out that at each formal order of perturbation theory, which is now 
reorganized by the presence of additional three-dimensional diagrams
due to the constraints, the leading infra-red terms are free of spurious
singularities and can be represented in three-dimensional $x$-space as
local functions of $r$, proportional to $(g^2/r)^n$, where $r$ is the c.m.
relative distance, $g$ the coupling constant and $n$ the formal order
of perturbation theory. The series of leading terms can be summed and result 
in local functions (in $r$) for the expressions of the potentials. The 
latter are compatible with the potentials proposed by Todorov in the
quasipotential approach on the basis of minimal substitution rules \cite{t2}
and later investigated in the fermionic case by Crater and Van Alstine
\cite{cva1}.
\par
The purpose of the present paper is to take into account, in the case
of vector interactions, the effects of the anomalous magnetic moments of
fermions. In usual calculations in QED, because of the smallness of the
coupling constant, the latter are evaluated at leading order of 
perturbation theory in the nonrelativistic limit ($O(\alpha^5)$ effect,
where $\alpha$ is the fine structure constant). However, in strong
coupling problems, like in the strong coupling regime of QED or in the
short-distance (vector) interactions of QCD, nonperturbative 
contributions of the anomalous magnetic moments may become sizable and
the incorporation of higher order effects becomes necessary.
\par
In order to include the main effects of the anomalous magnetic moments
into the potentials, we evaluate their contributions through the vertex
corrections. In the lowest order graph the anomalous magnetic moment
appears by means of a substitution rule that replaces each charge 
(coupling constant) by a Dirac-matrix function. We then introduce
this typical vertex correction at the vertices in each order of the
perturbation series of the vector potential determined previously 
\cite{js}.
Although the substitution rule utilized above is rather simple, it
generates in the higher order terms technical complications for the
summation of the perturbation series of the potential. The reason for
this is the new Dirac-matrix structure that results from the higher
order terms. Up to now, all the potentials that were considered in the 
constraint theory wave equations had 
dependences on the Dirac matrices only through pairs of $\gamma_1$ and
$\gamma_2$ matrices, the indices 1 and 2 referring to the two fermions,
respectively (like $\gamma_1.\gamma_2$, $\gamma_{15}\gamma_{25}$, etc.)$-$
a feature that considerably simplifies many algebraic operations as 
well as the reduction process to the final eigenvalue equation. The 
presence of the vertex corrections, even though globally symmetric in the
exchanges $1\leftrightarrow 2$, breaks this symmetry in the individual
terms and introduces new types of structure not present in previous
calculations. It is the presence of these terms that makes calculations
rather complicated. While the potential can still be represented in a
somehow compact form, the final eigenvalue equation for general quantum
numbers becomes less easy to handle. It takes a relatively simple form
only in the sectors of $j=0$ states (pseudoscalar and scalar), to which
we have limited our final analysis. The ground states of these sectors
are precisely those which may be concerned with spontaneous breakdowns
of symmetries (chiral and dilatational).
\par
The plan of the paper is as follows. In Sec. 2, we consider lowest order 
perturbation theory and determine the substitution rule to be used for the
vertex correction. In Sec. 3, the vertex corrections are incorporated into
the vector potential. The new form of the latter is determined in Sec. 4,
by resumming the corresponding perturbation series. In Sec. 5, the wave
equations are reduced, for the $j=0$ states, to a final eigenvalue 
equation. In Sec. 6, we analyze the effects of the anomalous magnetic
moments in several limiting cases. The requirement from the accompanying
form factors of not aggravating the singularities of the initial
potential, leads to restrictions on their behavior near the origin in 
$x$-space. The nonrelativistic as well as the one-particle limits are also
checked. Concluding remarks follow in Sec. 7.
\par

\section{Structure of the two-body potential with \protect \\ anomalous 
magnetic moments}
\setcounter{equation}{0}

In order to determine the structure of the potential in the presence of
anomalous magnetic moments, we start from the expression of the coupling
of a pointlike particle (fermion 1) with charge $e_1$ and anomalous
magnetic moment $\kappa_1$ to an external electromagnetic potential 
$A_{\mu}$ and to its field strength tensor $F_{\mu\nu}$:
\begin{equation} \label{2e1}
(\gamma_{1\mu} A^{\mu} + \frac{1}{2}\kappa_1 \sigma _{\mu\nu} 
F^{\mu\nu})
\end{equation}
[$\sigma_{\mu\nu}=\frac{1}{2i}[\gamma_{\mu},\gamma_{\nu}]$]. In the case
of a mutual interaction with another particle 2, expression
(\ref{2e1}) represents the lowest order perturbation theory result, 
where potential $A_{\mu}$ is itself expressed in terms of the photon 
propagator and its coupling to particle 2 (which we suppose for the moment 
without anomalous magnetic moment):
\begin{equation} \label{2e2}
A_{\mu}\ =\ D_{\mu\nu} \gamma_2^{\nu}\ ,
\end{equation}
where $D_{\mu\nu}$ is the photon propagator including the coupling
constants at its two ends.
\par
In the three-dimensional formalism of constraint theory, the 
Bethe-Salpeter kernel is projected on the constraint hypersurface and 
the wave function expanded around it \cite{s2,bb}. In the c.m. frame,
this amounts to projecting the kernel on the hypersurface where the
temporal component of the momentum transfer is zero. In a covariant
formalism, one first decomposes four-vectors along transverse and
longitudinal components with respect to the total momentum $P$:
\begin{eqnarray} \label{2e3}
& & P\ =\ p_1+p_2\ ,\ \ \ p\ =\ \frac {1}{2}(p_1-p_2)\ ,\ \ \ 
X\ =\ \frac {1}{2}(x_1+x_2)\ ,\ \ \ x\ =\ x_1-x_2\ ,\nonumber \\
& & x_{\mu}^T\ =\ x_{\mu}-\hat P.x \hat P_{\mu}\ ,\ \ \ x_L\  =\ 
\hat P.x\ ,\ \ \ P_L\ =\ \sqrt{P^2}\ ,\ \ \ \hat P_{\mu}\ =\ \frac
{P_{\mu}}{P_L}\ ,\nonumber \\
& & x^{T2}\ =\ x^2-(\hat P.x)^2\ ,\ \ \ r\ =\ \sqrt{-x^{T2}}\ ,
\nonumber \\
& & \gamma_{\mu}^T\ =\ \gamma _{\mu}-\hat P.\gamma \hat P_{\mu}\ ,
\ \ \ \gamma_L\ =\ \hat P.\gamma\ ,\ \ \ M\ =\ m_1+m_2\ .
\end{eqnarray}
Thus, in constraint theory, the propagator in Eq. (\ref{2e2}) depends
on $x$, in $x$-space, through $x^T$ only; in the Feynman gauge, to which
we stick throughout this work, it has the expression (in lowest order)
\cite{js}:
\begin{equation} \label{2e4}
D_{\mu\nu}\ =\ g_{\mu\nu} D(x^{T2},P_L)\ ,\ \ \ D(x^{T2},P_L)\ =\ 
\frac {e_1e_2}{4\pi} \frac {1}{2P_L r}\ .
\end{equation}
\par
Using Eqs. (\ref{2e2}) and (\ref{2e4}) the field strength tensor takes 
the form:
\begin{eqnarray} \label{2e5}
F_{\mu\nu} &=& \partial _{\mu} A_{\nu} - \partial _{\nu} A_{\mu}\ =\
2\dot D (x_{\mu}^T \gamma_{2\nu} - x_{\nu}^T \gamma _{2\mu})\nonumber \\
&=& \frac {1}{r^2} D (x_{\mu}^T \gamma _{2\nu} - x_{\nu}^T \gamma _{2\mu}) 
\ ,
\end{eqnarray}
where the dot operation represents derivation with respect to $x^{T2}$:
\begin{equation} \label{2e6}
\dot f\ \equiv \ \frac {\partial}{\partial x^{T2}} f\ .
\end{equation}
\par
Expression (\ref{2e1}) then becomes:
\begin{eqnarray} \label{2e7}
\gamma_{1\mu}A^{\mu} + \frac{1}{2}\kappa_1 \sigma _{1\mu\nu}F^{\mu\nu}
&=& D [ \gamma_1.\gamma_2 - i\frac{\kappa_1}{2r^2}(\gamma_1^T.x^T 
\gamma_1.\gamma_2 - \gamma_1.\gamma_2 \gamma_1^T.x^T) ]\nonumber \\
&=& \frac{1}{2} D (1-i\frac{\kappa_1}{r} \gamma_1^T.\frac{x^T}{r})
\gamma_1.\gamma_2\nonumber \\
& &\ \ \ + \gamma_{10}\gamma_{20} \big [ \frac {1}{2} D (1-i\frac
{\kappa_1}{r} \gamma_1^T.\frac{x^T}{r}) \gamma_1.\gamma_2 \big ]^{\dagger}
\gamma_{10}\gamma_{20}\ .
\end{eqnarray}
[The dagger represents hermitian conjugation.] We deduce that to this 
order the anomalous magnetic moment appears through the following matrix
substitution of the charge $e_1$:
\begin{equation} \label{2e8}
e_1\ \rightarrow \ e_1' = e_1(1-i\frac{\kappa_1}{r} \gamma_1^T.\frac
{x^T}{r})\ .
\end{equation} 
The above calculations can be repeated at the particle 2 (antifermion)
vertex, where the charge substitution becomes:
\begin{equation} \label{2e9}
e_2\ \rightarrow \ e_2' = e_2(1-i\frac{\kappa_2}{r} \gamma_2^T.\frac
{x^T}{r})\ .
\end{equation}
[$e_2$ is the fermion 2 charge; the passage to the antifermion is obtained 
in momentum space with the replacement $p_1\rightarrow -p_2$, which yields 
for the momentum transfer $q=p_1-p_1'\rightarrow -p_2+p_2'=q$ and hence in 
$x$-space $x\rightarrow x$; the Dirac matrices $\gamma_2$ act on the
wave function from the right.]
\par
The mutual interaction potential then becomes to lowest order:
\begin{eqnarray} \label{2e10}
V &=& \frac{1}{2} D (1-i\frac{\kappa_1}{r} \gamma_1^T.\frac{x^T}{r})
(1-i\frac{\kappa_2}{r} \gamma_2^T.\frac{x^T}{r}) \gamma_1.\gamma_2
\nonumber \\
& &\ \ +\gamma_{10}\gamma_{20} \big [ \frac{1}{2} D (1-i\frac{\kappa_1}{r}
\gamma_1^T.\frac{x^T}{r}) (1-i\frac{\kappa_2}{r} \gamma_2^T.\frac{x^T}{r})
\gamma_1.\gamma_2 \big ]^{\dagger} \gamma_{10}\gamma_{20}\ .
\end{eqnarray}
\par
It is natural to generalize the substitution rules (\ref{2e8})-(\ref{2e9})
to higher orders, with the difference that the lowest order anomalous
magnetic term $\kappa/r$ should now be replaced by a form factor $b(r)$:
\begin{equation} \label{2e11}
e_a\ \rightarrow\ e_a'=e_a \big (1-ib(r)\gamma_a^T.\frac{x^T}{r}\big )
\ \ \ \ \ (a=1,2)\ .
\end{equation}
The form factor $b(r)$ should ensure for the anomalous magnetic term
a smooth behavior at the origin, in order not to enhance the existing
singularities of the propagator $D$. Taking into account the lowest order
expression of $\kappa$, $\kappa = \frac{\hbar}{2m}\frac{\alpha}{2\pi}$,
a convenient parametrization for $b(r)$ is:
\begin{eqnarray} \label{2e12}
b_a(r) &=& \frac{\hbar \alpha/(2\pi)}{2mr+(\hbar \alpha/(2\pi)) c_a(r)}\ ,
\ \ \ \ c_a(r)>1\ \ \ \ (a=1,2)\ ,\nonumber \\
\alpha &=& \frac {e^2}{4\pi}\ ,\ \ \ e_1 = e_2 = e\ .
\end{eqnarray}
\par
Expression (\ref{2e10}) is then generalized to the following form:
\begin{eqnarray} \label{2e13}
V &=& \frac{1}{2}\big (A-iB_1\gamma_1^T.\frac{x^T}{r}-iB_2\gamma_2^T.
\frac{x^T}{r}-C\gamma_1^T.\frac{x^T}{r}\gamma_2^T.\frac{x^T}{r}\big )
\gamma_1.\gamma_2\nonumber \\
& &\ \ \ +\frac{1}{2}\gamma_1.\gamma_2\big (A+iB_1\gamma_1^T.\frac{x^T}
{r}+iB_2\gamma_2^T.\frac{x^T}{r}-C\gamma_1^T.\frac{x^T}{r}\gamma_2^T.
\frac{x^T}{r}\big )\ ,
\end{eqnarray}
where the potentials $A$, $B_1$, $B_2$ and $C$ are completely determined,
by means of the substitution rules (\ref{2e11}), from the expression of 
$V$ in the absence of anomalous magnetic moments. The latter expression
has the form:
\begin{equation} \label{2e14}
V_0\ =\ A_0 \gamma_1.\gamma_2\ .
\end{equation}
[$A_0$ is denoted by  $V_2$ in Ref. \cite{js}.]
\par
For the Todorov potential \cite{t2,cva1,js}, $A_0$ is:
\begin{equation} \label{2e15}
A_0\ =\ \frac{1}{4} \ln (1+\frac{2\alpha}{P_L r})\ .
\end{equation}
However, other effective expressions could be used for $A_0$ as well.
\par
Actually, the potentials that appear in the constraint theory wave 
equations are functions of $V$ [Eq. (\ref{2e13})] through exponentiations;
therefore,  we shall need to calculate such exponential functions. This
is the main content of Sec. 3.
\par

\newpage

\section{Wave equations}
\setcounter{equation}{0}

The wave equations of constraint theory for a fermion-antifermion system
can be written in the form \cite{s1,ms1}:
\begin{eqnarray} \label{3e1}
(\gamma_1.p_1-m_1)\ \widetilde \Psi &=& (-\gamma_2.p_2+m_2) \widetilde V\
\widetilde \Psi\ ,\nonumber \\
(-\gamma_2.p_2-m_2)\ \widetilde \Psi &=& (\gamma_1.p_1+m_1) \widetilde V\
\widetilde \Psi\ ,
\end{eqnarray}
where $\widetilde \Psi$ is a spinor function of rank two, represented as a
$4\times 4$ matrix function; the Dirac matrices $(\gamma_2)$ of the 
antifermion act on $\widetilde \Psi$ from the right; the total and 
relative variables were defined in Eqs. (\ref{2e3}); $\widetilde V$ is a
Poincar\'e invariant mutual interaction potential.
\par
Equations (\ref{3e1}) imply the constraint
\begin{equation} \label{3e2}
\big [ (p_1^2-p_2^2)-(m_1^2-m_2^2)\big ]\ \widetilde \Psi\ =\ 0\ ,
\end{equation}
or equivalently
\begin{equation} \label{3e3}
C(p)\ \equiv\ 2P_Lp_L-(m_1^2-m_2^2)\ \simeq\ 0\ ,
\end{equation}
which allows the elimination from the wave equations of the relative 
longitudinal momentum in terms of the masses and the c.m. total energy.
The wave function $\widetilde \Psi$, for eigenfunctions of the total
momentum $P$, has then the structure:
\begin{equation} \label{3e4}
\widetilde \Psi (X,x)\ =\ e^{{\displaystyle -iP.X}} e^{{\displaystyle
-i(m_1^2-m_2^2)x_L/(2P_L)}}\ \widetilde \psi(x^T)\ .
\end{equation}
\par
The positivity conditions of the norm of $\widetilde \Psi$ imply that
$\widetilde V$ should satisfy the inequality $\frac{1}{4} Tr \widetilde
V^{\dagger}\widetilde V < 1$ \cite{s1,ms1}. A convenient parametrization
satisfying this inequality for potentials commuting with $\gamma_{1L}
\gamma_{2L}$ was proposed by Crater and Van Alstine \cite{cva2}; it is:
$\widetilde V = \tanh V$. It turns out that the perturbation series of
the leading infra-red terms in QED in the Feynman gauge provides a
potential $V$ that is compatible with this parametrization \cite{js}
(cf. Eqs. (\ref{2e14})-(\ref{2e15}), where $V$ is denoted by $V_0$).
For more general potentials which do not commute with $\gamma_{1L}
\gamma_{2L}$, the generalization of the above parametrization is 
\cite{ms1}:
\begin{equation} \label{3e5}
\gamma_{1L}\gamma_{2L} \widetilde V\ =\ \tanh (\gamma_{1L}\gamma_{2L}
V)\ .
\end{equation}
Equations (\ref{3e1}) are then transformed with the change of wave
function:
\begin{equation} \label{3e6}
\widetilde \Psi\ =\ \cosh (\gamma_{1L}\gamma_{2L} V)\ \Psi\ ;
\end{equation}
they become:
\begin{eqnarray} \label{3e7}
(\gamma_1.p_1-m_1)\cosh (\gamma_{1L}\gamma_{2L}V)\ \Psi &=& (-\gamma_2.
p_2+m_2)\gamma_{1L}\gamma_{2L}\sinh (\gamma_{1L}\gamma_{2L}V)\ \Psi\ ,
\nonumber \\
(-\gamma_2.p_2-m_2)\cosh (\gamma_{1L}\gamma_{2L}V)\ \Psi &=& (\gamma_1.
p_1+m_1)\gamma_{1L}\gamma_{2L}\sinh (\gamma_{1L}\gamma_{2L}V)\ \Psi\ .
\end{eqnarray}
\par
One can also equivalently work in the ``Breit representation''. Defining
\begin{eqnarray}
\label{3e8}
V_B &=& \gamma_{1L}\gamma_{2L} V\ ,\\
\label{3e9}
\Psi_B &=& e^{{\displaystyle -V_B}} \Psi\ ,
\end{eqnarray}
one shows that Eqs. (\ref{3e1}) or (\ref{3e7}) reduce to thye Breit type
equation \cite{ms1}:
\begin{equation} \label{3e10}
\big [ P_L e^{{\displaystyle 2V_B}} - ({\cal H}_1+{\cal H}_2)\big ]\
\Psi\ =\ 0\ ,
\end{equation}
provided constraint (\ref{3e2})-(\ref{3e3}) is used; here, ${\cal H}_1$
and ${\cal H}_2$ are the covariant free hamiltonians:
\begin{eqnarray} \label{3e11}
{\cal H}_1 &=& m_1\gamma_{1L} - \gamma_{1L}\gamma_1^T.p_1^T\ ,\nonumber \\
{\cal H}_2 &=& -m_2\gamma_{2L} - \gamma_{2L}\gamma_2^T.p_2^T\ .
\end{eqnarray}
\par
The normalization conditions of the wave functions $\widetilde \Psi$, 
$\Psi$ and $\Psi_B$ were presented in Ref. \cite{ms1}.
\par
To solve the wave equations one decomposes the sixteen-component 
($4\times 4$) wave function $\psi$ along four-component ($2\times 2$) wave
functions:
\begin{equation} \label{3e12}
\psi\ =\ \psi_1+\gamma_L\psi_2+\gamma_5\psi_3+\gamma_L\gamma_5\psi_4\ ,
\end{equation}
and similarly for the Breit type wave function $\psi_B$:
\begin{equation} \label{3e13}
\psi_B\ =\ \psi_{B1}+\gamma_L\psi_{B2}+\gamma_5\psi_{B3}+\gamma_L\gamma_5
\psi_{B4}\ .
\end{equation}
These components are obtained with the projectors \cite{ms1}
\begin{eqnarray} \label{3e14}
& &{\cal P}_1\ =\ \frac{1}{4}(1+\gamma_{1L}\gamma_{2L})(1+\gamma_{15}
\gamma_{25})\ ,\ \ \ \ {\cal P}_2\ =\ \frac{1}{4}(1+\gamma_{1L}
\gamma_{2L})(1-\gamma_{15}\gamma_{25})\ ,\nonumber \\
& &{\cal P}_3\ =\ \frac{1}{4}(1-\gamma_{1L}\gamma_{2L})(1+\gamma_{15}
\gamma_{25})\ ,\ \ \ \ {\cal P}_4\ =\ \frac{1}{4}(1-\gamma_{1L}
\gamma_{2L})(1-\gamma_{15}\gamma_{25})\ .
\end{eqnarray}
\par
The spin operators, which act in the four-component wave function
subspaces, are defined by means of the Pauli-Lubanski operators:
\begin{eqnarray} \label{3e15}
& & W_{1S\alpha}\ =\ -\frac{\hbar}{4}\epsilon_{\alpha\beta\mu\nu}
P^{\beta}\sigma_1^{\mu\nu}\ ,\ \ \ \ W_{2S\alpha}\ =\ -\frac{1}{4}
\epsilon_{\alpha\beta\mu\nu}P^{\beta}\sigma_2^{\mu\nu}\ \ \ \ \
(\epsilon_{0123}=+1)\ ,\nonumber \\
& & \gamma_{1L}W_{1S\alpha}\ =\ \frac{\hbar P_L}{2} \gamma_{1\alpha}^T
\gamma_{15}\ ,\ \ \ \ \gamma_{2L}W_{2S\alpha}\ =\ \frac{\hbar P_L}{2}
\gamma_{2\alpha}^T\gamma_{25}\ ,\nonumber \\
& & W_{1S}^2\ =\ W_{2S}^2\ =\ -\frac{3}{4}\hbar^2 P^2\ ,\ \ \ \ 
W_S\ =\ W_{1S}+W_{2S}\ ,\nonumber \\
& & w\ \equiv \ \big (\frac{2}{\hbar P_L}\big )^2 W_{1S}.W_{2S}\
_{\stackrel {{\displaystyle \longrightarrow}}{{\rm c.m.}}}\ -\frac{4}
{\hbar^2} {\bf s}_1.{\bf s}_2\ ,\nonumber \\
& & w_{12}\ \equiv \ \big (\frac{2}{\hbar P_L}\big )^2\frac {W_{1S}.x^T
W_{2S}.x^T}{x^{T2}}\ _{\stackrel {{\displaystyle \longrightarrow}}
{{\rm c.m.}}}\ -\frac{4}{\hbar^2}\frac{({\bf s}_1.{\bf x})({\bf s}_2.
{\bf x})}{{\bf x}^2}\ ,\nonumber \\
& & w_{12}^2\ =\ 1\ ,\ \ \ \ w_{12}(w-w_{12})\ =\ w-w_{12}\ .
\end{eqnarray}
\par
It is clear, from Eqs. (\ref{3e7}) and (\ref{3e10}), that one has to 
calculate the exponential of $\gamma_{1L}\gamma_{2L}V$, with
$V$ having the general structure (\ref{2e13}). We have: 
\begin{eqnarray} \label{3e16}
V_B &=& \gamma_{1L}\gamma_{2L}V\nonumber \\
&=& \frac{1}{2}\big (A+iB_1\gamma_1^T.\frac{x^T}{r}+iB_2\gamma_2^T.
\frac{x^T}{r}-C\gamma_1.\frac{x^T}{r}\gamma_2.\frac{x^T}{r}\big )\
\gamma_{1L}\gamma_{2L}\gamma_1.\gamma_2\nonumber \\
& &\ +\ \frac{1}{2}\gamma_{1L}\gamma_{2L}\gamma_1.\gamma_2\ \big (
A+iB_1\gamma_1^T.\frac{x^T}{r}+iB_2\gamma_2^T.\frac{x^T}{r}-C
\gamma_1^T.\frac{x^T}{r}\gamma_2^T.\frac{x^T}{r}\big )\ .
\end{eqnarray}
The difficulty of the calculation stems from the fact that the matrices
$\gamma_1^T.x^T$ and $\gamma_2^T.x^T$ do not commute with $\gamma_1.
\gamma_2$.
\par
To proceed further, we introduce in the subspace of $x^T$ a longitudinal
direction, parallel to $x^T$, and a transverse plane, orthogonal to it.
We define:
\begin{eqnarray} \label{3e17}
& & \hat x_{\mu}^T\ =\ \frac{x_{\mu}^T}{\sqrt{-x^{T2}}}\ =\ \frac{x_{\mu}
^T}{r}\ ,\ \ \ \ \hat x^{T2}\ =\ -1\ ,\nonumber \\
& & \gamma_{a\mu}^T\ =\ \gamma_{a\mu}^{\ell}+\gamma_{a\mu}^t\ =\ 
-\gamma_a.\hat x^T \hat x_{\mu}^T+\gamma_{a\mu}^t\ ,\ \ \ \ \gamma_a^t.
\hat x^T\ =\ 0\ ,\nonumber \\
& & \gamma_{a\ell}\ =\ \gamma_a^T.\hat x^T\ ,\ \ \ \ \gamma_{a\ell}^2\
=\ -1\ \ \ \ \ (a=1,2)\ .
\end{eqnarray}
The capital indices $L$ and $T$ [Eqs. (\ref{2e3})] concern the 
longitudinal and transverse components with respect to the total
momentum $P$, while the small indices $\ell$ and $t$ concern those of 
the three-dimensional relative distance $x^T$. We list here some useful
relations satisfied by these matrices:
\begin{eqnarray} \label{3e18}
& & \gamma_1.\gamma_2\ =\ \gamma_{1L}\gamma_{2L}+\gamma_1^T.\gamma_2^T\
=\ \gamma_{1L}\gamma_{2L}-\gamma_{1\ell}\gamma_{2\ell}+\gamma_1^t.\gamma
_2^t\ ,\nonumber \\
& & \gamma_{1\ell}\gamma_{2\ell}\ =\ -\gamma_{1L}\gamma_{2L}\gamma_{15}
\gamma_{25}w_{12}\ ,\ \ \ \ \gamma_1^T.\gamma_2^T\ =\ -\gamma_{1L}
\gamma_{2L}\gamma_{15}\gamma_{25}w\ ,\nonumber \\  
& & \gamma_1^t.\gamma_2^t\ =\ -\gamma_{1\ell}\gamma_{2\ell}(w-w_{12})\ ,
\ \ \ \ \big [\gamma_{a\ell}, \gamma_{aL}\big ]_+\ =\ 0\ ,\nonumber \\
& & \big [\gamma_{a\ell},(w-w_{12})\big ]_+\ =\ 0\ ,\ \ \ \ 
\big [\gamma_{a\ell},\gamma_{aL}(w-w_{12})\big ]\ =\ 0\ \ \ \ (a=1,2)\ .
\end{eqnarray}
($[\ ,\ ]_+$ is the anticommutator.)
\par
At the first stage of the calculation, one can eliminate $\gamma_{a5}$
and $\gamma_a^t$ ($a=1,2$) in terms of $\gamma_{a\ell}$, $\gamma_{aL}$,
$w$ and $w_{12}$. For the subspace of the matrices $\gamma_{a\ell}$, one
introduces the following projectors:
\begin{eqnarray} \label{3e19}
& & {\cal P}_{++}\ =\ \frac{1}{4}(1+i\gamma_{1\ell})(1+i\gamma_{2\ell})\ ,
\ \ \ \ {\cal P}_{+-}\ =\ \frac{1}{4}(1+i\gamma_{1\ell})(1-i\gamma_{2\ell})
\ ,\nonumber \\
& & {\cal P}_{-+}\ =\ \frac{1}{4}(1-i\gamma_{1\ell})(1+i\gamma_{2\ell})\ ,
\ \ \ \ {\cal P}_{--}\ =\ \frac{1}{4}(1-i\gamma_{1\ell})(1-i\gamma_{2\ell})
\ ,
\end{eqnarray}
which allows the decomposition of the $\gamma_{a\ell}$'s along the latter.
Potential $V_B$ [Eq. (\ref{3e16})] takes now the form:
\begin{eqnarray} \label{3e20}
V_B\ =\ \gamma_{1L}\gamma_{2L}V &=& -(A\gamma_{1\ell}\gamma_{2\ell}-C)
\gamma_{1L}\gamma_{2L}+(A-C\gamma_{1\ell}\gamma_{2\ell})(1+\gamma_{1L}
\gamma_{2L}\gamma_1^t.\gamma_2^t)\nonumber \\
& &\ +i(B_1\gamma_{1\ell}+B_2\gamma_{2\ell})(1+\gamma_{1L}\gamma_{2L}
\gamma_1^t.\gamma_2^t)\ .
\end{eqnarray}
Notice that the second term in the right-hand side above commutes with
the two others and therefore its exponential can be factorized and 
calculated independently. The first and third terms can be written in 
terms of the projectors (\ref{3e19}):
\begin{eqnarray} \label{3e21}
& & -A\gamma_{1\ell}\gamma_{2\ell}\gamma_{1L}\gamma_{2L}+C\gamma_{1L}
\gamma_{2L} + i(B_1\gamma_{1\ell}+B_2\gamma_{2\ell})(1+\gamma_{1L}
\gamma_{2L}\gamma_1^t.\gamma_2^t)\nonumber \\
& &\ \ \ \ =\ {\cal P}_{++}\big [(A+C)\gamma_{1L}\gamma_{2L}+(B_1+B_2)
(1+\gamma_{1L}\gamma_{2L}(w-w_{12}))\big ]\nonumber \\
& &\ \ \ \ \ \ +{\cal P}_{+-}\big [(-A+C)\gamma_{1L}\gamma_{2L}+(B_1-B_2)
(1-\gamma_{1L}\gamma_{2L}(w-w_{12}))\big ]\nonumber \\
& &\ \ \ \ \ \ +{\cal P}_{-+}\big [(-A+C)\gamma_{1L}\gamma_{2L}-(B_1-B_2)
(1-\gamma_{1L}\gamma_{2L}(w-w_{12}))\big ]\nonumber \\
& &\ \ \ \ \ \ +{\cal P}_{--}\big [(A+C)\gamma_{1L}\gamma_{2L}-(B_1+B_2)
(1+\gamma_{1L}\gamma_{2L}(w-w_{12}))\big ]\ .
\end{eqnarray}
\par
The exponential of this expression can be calculated by a series
expansion. The projectors ${\cal P}_{++}$, etc., commute with 
$\gamma_{1L}\gamma_{2L}(w-w_{12})$, but satisfy particular commutation
rules with $\gamma_{1L}\gamma_{2L}$. One factorizes the projectors 
${\cal P}_{++}$, etc., on the left of the series. Each multiplicative 
factor of ${\cal P}_{++}$, etc., can be resummed into exponential 
functions. At the end, one rexpresses $\gamma_{1\ell}$ and 
$\gamma_{2\ell}$ in terms of $\gamma_{1L}$, $\gamma_{2L}$, $\gamma_{15}$,
$\gamma_{25}$ and the spin operators (\ref{3e15}) and one introduces back
the projectors ${\cal P}_i$ ($i=1,\ldots ,4$), (\ref{3e14}). One thus
obtains:
\newpage
\begin{eqnarray} \label{3e22}
e^{{\displaystyle 2V_B}} &=& e^{{\displaystyle 2\gamma_{1L}\gamma_{2L}
V}}\nonumber \\
&=&\ \frac{1}{2}(1+w_{12})e^{{\displaystyle \alpha_+\gamma_+}} (f_{++}
+g_{++}){\cal P}_1 + \frac{1}{2}(1-w_{12})e^{{\displaystyle \alpha_-
\gamma_+}} (f_{--}-g_{--}){\cal P}_1\nonumber \\
& & +\frac{1}{2}(1-w_{12})e^{{\displaystyle \alpha_+\gamma_-}} (f_{++}
+g_{++}){\cal P}_2 + \frac{1}{2}(1+w_{12})e^{{\displaystyle \alpha_-
\gamma_-}} (f_{--}-g_{--}){\cal P}_2\nonumber \\
& & +\frac{1}{2}(1-w_{12})e^{{\displaystyle \alpha_+\gamma_+}} (f_{+-}
-g_{+-}){\cal P}_3 + \frac{1}{2}(1+w_{12})e^{{\displaystyle \alpha_-
\gamma_+}} (f_{-+}+g_{-+}){\cal P}_3\nonumber \\
& & +\frac{1}{2}(1+w_{12})e^{{\displaystyle \alpha_+\gamma_-}} (f_{+-}
-g_{+-}){\cal P}_4 + \frac{1}{2}(1-w_{12})e^{{\displaystyle \alpha_-
\gamma_-}} (f_{-+}+g_{-+}){\cal P}_4\nonumber \\
& & +\frac{i}{2}(\gamma_{1\ell}+\gamma_{2\ell})\big [e^{{\displaystyle
\alpha_+\gamma_+}}h_{++}{\cal P}_1 + e^{{\displaystyle \alpha_+\gamma_-}}
h_{++}{\cal P}_2 + e^{{\displaystyle \alpha_+\gamma_+}}h_{+-}{\cal P}_3
+ e^{{\displaystyle \alpha_+\gamma_-}}h_{+-}{\cal P}_4\big ]\nonumber \\
& & +\frac{i}{2}(\gamma_{1\ell}-\gamma_{2\ell})\big [e^{{\displaystyle
\alpha_-\gamma_+}}h_{--}{\cal P}_1 + e^{{\displaystyle \alpha_-\gamma_-}}
h_{--}{\cal P}_2 + e^{{\displaystyle \alpha_-\gamma_+}}h_{-+}{\cal P}_3
+ e^{{\displaystyle \alpha_-\gamma_-}}h_{-+}{\cal P}_4\big ]\ .\nonumber \\
& &
\end{eqnarray}
The definitions of the potential functions are the following:
\begin{eqnarray} 
\label{3e23}
& & f_{rs}\ =\ \cosh \sqrt{\alpha_r^2+\beta_r^2\gamma_s^2}\ ,\ \ \ \
g_{rs}\ =\ \frac{\alpha_r}{\sqrt{\alpha_r^2+\beta_r^2\gamma_s^2}}\sinh
\sqrt{\alpha_r^2+\beta_r^2\gamma_s^2}\ ,\nonumber \\
& & h_{rs}\ =\ \frac{\beta_r\gamma_s}{\sqrt{\alpha_r^2+\beta_r^2
\gamma_s^2}}\sinh \sqrt{\alpha_r^2+\beta_r^2\gamma_s^2}\ ,\ \ \ \ 
r,s=\pm\ ,\nonumber \\
& & f_{rs}^2-g_{rs}^2-h_{rs}^2\ =\ 1\ ,\\
\label{3e24}
& & \alpha_{\pm}\ =\ 2(A\pm C)\ ,\ \ \ \ \beta_{\pm}\ =\ 2(B_1\pm B_2)\ ,
\nonumber \\
& & \gamma_{\pm}\ =\ 1\pm (w-w_{12})\ .
\end{eqnarray}
The exponential function $e^{{\displaystyle -2V_B}}$ is obtained from Eq.
(\ref{3e22}) by the replacements $\alpha \rightarrow -\alpha$ and $\beta
\rightarrow -\beta$; $e^{{\displaystyle V_B}}$ is obtained by the 
replacements $\alpha \rightarrow \alpha/2$ and $\beta \rightarrow \beta/2$,
etc.. Also notice the commutation relations:
\begin{eqnarray} \label{3e25}
& & \gamma_{a\ell}\gamma_{\pm}\ =\ \gamma_{\mp}\gamma_{a\ell}\ ,\ \ \ \ 
\gamma_{a\ell}f_{\pm\pm}\ =\ f_{\pm\mp}\gamma_{a\ell}\ ,\nonumber \\
& & \gamma_{a\ell}g_{\pm\pm}\ =\ g_{\pm\mp}\gamma_{a\ell}\ ,\ \ \ \ 
\gamma_{a\ell}h_{\pm\pm}\ =\ h_{\pm\mp}\gamma_{a\ell}\ ,\ \ \ \ a=1,2\ .
\end{eqnarray}
\par
Equation (\ref{3e22}) and the similar ones with different arguments allow
us to project the wave equations (\ref{3e7}) or (\ref{3e10}) with the 
aid of the projectors ${\cal P}_i$ ($i=1,\ldots ,4$), (\ref{3e14}),
appearing on the utmost right of the expressions, on the four-component
wave functions (\ref{3e12}) or (\ref{3e13}). One thus obtains coupled
equations for the four components $\psi_i$ or $\psi_{Bi}$ ($i=1,\ldots,
4$) and eliminating three of them one reaches a final eigenvalue 
equation involving only one of the components.
\par

\section{Determination of the potentials}
\setcounter{equation}{0}

Before proceeding to the reduction of the wave equations, we shall determine
the expressions of the various potentials appearing in Eqs. (\ref{3e16})
and (\ref{3e22})-(\ref{3e24}) in terms of the elementary Coulomb potential.
\par
Potential $V$ has the structure (\ref{2e13}) and is obtained from Eq.
(\ref{2e14}) with the substitutions (\ref{2e11}). We shall, for the 
moment, not use the particular expression (\ref{2e15}) of the initial
potential $A_0$, but rather present the calculations for the general case.
We assume that $A_0$ is expressible as a power series of the elementary
Coulomb potential $\alpha/(2P_L r)$:
\begin{eqnarray} 
\label{4e1}
& & A_0\ =\ A_0(v)\ =\ \sum_{n=1}^{\infty} a_nv^n\ ,\\
\label{4e2}
& & v\ =\ \frac{\alpha}{2P_L r}\ .
\end{eqnarray}
\par
The substitutions (\ref{2e11}) yield for the fine structure constant 
$\alpha$ the modification:
\begin{equation} \label{4e3}
\alpha\ \longrightarrow \ \alpha^{\prime} = \alpha (1-ib_1\gamma_{1\ell})
(1-ib_2\gamma_{2\ell})\ ,
\end{equation}
which can be expressed in terms of the projectors (\ref{3e19}):
\begin{eqnarray} \label{4e4}
\alpha^{\prime} &=&\ \alpha \big [ (1-b_1)(1-b_2){\cal P}_{++} +
(1-b_1)(1+b_2){\cal P}_{+-}\nonumber \\
& &\ \ \ \ \ + (1+b_1)(1-b_2){\cal P}_{-+} + (1+b_1)(1+b_2){\cal P}_{--}\big ]\ .
\end{eqnarray}
Since $v$ [Eq. (\ref{4e2})] is proportional to $\alpha$, its 
modification is similar to that given by Eq. (\ref{4e4}). The initial
potential $A_0$ thus undergoes the change:
\begin{eqnarray} \label{4e5}
A_0\ \rightarrow \ \overline A_0 &=&\ \sum_{n=1}^{\infty} a_n v^n \big [
{\cal P}_{++}(1-b_1)^n(1-b_2)^n + {\cal P}_{+-}(1-b_1)^n(1+b_2)^n
\nonumber \\
& &\ \ \ \ \  + {\cal P}_{-+}(1+b_1)^n(1-b_2)^n + {\cal P}_{--}(1+b_1)^n(1+b_2)^n
\big ]\nonumber \\
&=&\ {\cal P}_{++}A_0\big (v(1-b_1)(1-b_2)\big ) + {\cal P}_{+-}A_0
\big (v(1-b_1)(1+b_2)\big )\nonumber \\
& & + {\cal P}_{-+}A_0\big (v(1+b_1)(1-b_2)\big ) + {\cal P}_{--}A_0\big (
v(1+b_1)(1+b_2)\big )\nonumber \\
&\equiv &\ {\cal P}_{++}V_{--} + {\cal P}_{+-}V_{-+} + {\cal P}_{-+}V_{+-}
+ {\cal P}_{--}V_{++}\ . 
\end{eqnarray}
Reexpressing the projectors ${\cal P}_{++}$, etc., in terms of the 
matrices $\gamma_{a\ell}$ ($a=1,2$), one obtains:
\begin{eqnarray} \label{4e6} 
\overline A_0 &=&\ \frac{1}{4}[V_{--}+V_{-+}+V_{+-}+V_{++}] +
\frac{1}{4}[V_{--}+V_{-+}-V_{+-}-V_{++}]i\gamma_{1\ell}\nonumber \\
& & + \frac{1}{4}[V_{--}-V_{-+}+V_{+-}-V_{++}]i\gamma_{2\ell} + 
\frac{1}{4}[V_{--}-V_{-+}-V_{+-}+V_{--}]i^2\gamma_{1\ell}\gamma_{2\ell}\ .
\nonumber \\
& &
\end{eqnarray}
This expression should be identified with the combination
$(A+iB_1\gamma_{1\ell}+iB_2\gamma_{2\ell}-C\gamma_{1\ell}\gamma_{2\ell})$
appearing in Eq. (\ref{3e16}) on the left or the right of $\gamma_{1L}
\gamma_{2L}\gamma_1.\gamma_2$ (the substitution (\ref{4e4}) being done
symmetrically with respect to this operator). We then obtain the
identifications:
\begin{eqnarray} \label{4e7}
& & A\ =\ \frac{1}{4}[V_{--}+V_{-+}+V_{+-}+V_{++}]\ ,\nonumber \\
& & B_1\ =\ \frac{1}{4}[V_{--}+V_{-+}-V_{+-}-V_{++}]\ ,\nonumber \\
& & B_2\ =\ \frac{1}{4}[V_{--}-V_{-+}+V_{+-}-V_{++}]\ ,\nonumber \\
& & C\ =\ \frac{1}{4}[V_{--}-V_{-+}-V_{+-}+V_{++}]\ ,
\end{eqnarray}
and from Eqs. (\ref{3e24}:
\begin{eqnarray} \label{4e8}
& & \alpha_+\ =\ V_{--}+V_{++}\ ,\ \ \ \ \alpha_-\ =\ V_{-+}+V_{+-}\ ,
\nonumber \\
& & \beta_+\ =\ V_{--}-V_{++}\ ,\ \ \ \ \beta_-\ =\ V_{-+}-V_{+-}\ .
\end{eqnarray}
\par
For the particular case of Todorov's potential (\ref{2e15}), the 
expressions of the potentials $V_{++}$, etc., are:
\begin{eqnarray} \label{4e9}
& & V_{--}\ =\ \frac{1}{4}\ln\big (1+4v(1-b_1)(1-b_2)\big )\ ,\nonumber \\
& & V_{-+}\ =\ \frac{1}{4}\ln\big (1+4v(1-b_1)(1+b_2)\big )\ ,\nonumber \\
& & V_{+-}\ =\ \frac{1}{4}\ln\big (1+4v(1+b_1)(1-b_2)\big )\ ,\nonumber \\
& & V_{++}\ =\ \frac{1}{4}\ln\big (1+4v(1+b_1)(1+b_2)\big )\ .
\end{eqnarray}
\par

\newpage

\section{Reduction to a final eigenvalue equation}
\setcounter{equation}{0}

In the absence of anomalous magnetic moments, the wave equations 
(\ref{3e7}) or (\ref{3e10}) can be reduced to a single 
Pauli-Schr\"odinger type equation for the component $\psi_3$ or 
$\psi_{B3}$ \cite{ms1}. A similar reduction can also be undertaken here;
however, due to the complexity of the new terms in the effective 
potential [Eq. (\ref{3e22})], the reduction process is not as
straightforward as before. The reason is that the components $\psi_i$
or $\psi_{Bi}$ ($i=1,\ldots ,4$) [Eqs. (\ref{3e12})-(\ref{3e13})] do no
longer have, in the general case, simple characterizations with the
quantum numbers $\ell$ (orbital angular momentum) and $s$ (total spin).
For instance, in the absence of anomalous magnetic moments, the component
$\psi_3$ can be classified according to the quantum numbers $\ell=j\pm 1$,
$s=1$ ($j$ being the total angular momentum) and $\ell = j$. In the 
present case, this simple property is lost and such a classification will
concern combinations of $\psi_3$ and $\psi_2$.
\par
It turns out that the most convenient representation where the reduction
process can be achieved is the ``anti-Breit'' representation defined with
the wave function transformations $\chi = e^{{\displaystyle V_B}}\psi =
e^{{\displaystyle 2V_B}}\psi_B$ [cf. Eq. (\ref{3e9})]. In this case the
reduced wave function is a tractable combination of $\chi_3$ and $\chi_2$.
We shall not, however, present here the reduced wave equation in the 
general cases of quantum numbers, the corresponding expression being
still lengthy, but rather shall content ourselves with the simplest case
of the $j=0$ quantum number, corresponding to the two sectors of 
pseudoscalar and scalar states. These are also the most sensitive sectors
involved in zero-mass bound state problems in strong coupling regimes.
\par
Actually, for these sectors, the Breit representation (\ref{3e9}) is the
simplest one and it is sufficient to project Eq. (\ref{3e10}) along the
components $\psi_{Bi}$ ($i=1,\ldots ,4$) [Eq. (\ref{3e13})]. In these
sectors, the operators $w_{12}$, $(W_{1S}+W_{2S}).\hat x^T$ and
$(W_{1S}+W_{2S}).p^T$ have the following quantum numbers: $w_{12}=1$,
$(W_{1S}+W_{2S}).\hat x^T = 0$, $(W_{1S}+W_{2S}).p^T = 0$. Equation
(\ref{3e10}), together with Eq. (\ref{3e22}), then yields the following
four coupled equations:
\begin{eqnarray} \label{5e1}
P_Le^{{\displaystyle \alpha_+\gamma_+}}(f_{++}+g_{++})\psi_{B1} &-&
(m_1-m_2)\psi_{B2}+(\frac{2}{\hbar P_L})(W_{1S}-W_{2S}).p^T\psi_{B3}
\nonumber \\
&-& \frac{i}{2}P_L(\frac{2}{\hbar P_L})(W_{1S}-W_{2S}).\hat x^T
e^{{\displaystyle \alpha_+\gamma_-}}h_{+-}\psi_{B4}\ =\ 0\ ,
\nonumber \\
P_Le^{{\displaystyle \alpha_-\gamma_-}}(f_{--}-g_{--})\psi_{B2} &-&
(m_1-m_2)\psi_{B1}\nonumber \\
&+& \frac{i}{2}P_L(\frac{2}{\hbar P_L})(W_{1S}-W_{2S}).\hat x^T
e^{{\displaystyle \alpha_-\gamma_+}}h_{-+}\psi_{B3}\ =\ 0\ ,
\nonumber \\
P_Le^{{\displaystyle \alpha_-\gamma_+}}(f_{-+}+g_{-+})\psi_{B3} &-&
M\psi_{B4}+(\frac{2}{\hbar P_L})(W_{1S}-W_{2S}).p^T\psi_{B1}
\nonumber \\
&-&\frac{i}{2}P_L(\frac{2}{\hbar P_L})(W_{1S}-W_{2S}).\hat x^T
e^{{\displaystyle \alpha_-\gamma_-}}h_{--}\psi_{B2}\ =\ 0\ ,
\nonumber \\
P_Le^{{\displaystyle \alpha_+\gamma_-}}(f_{+-}-g_{+-})\psi_{B4} &-&
M\psi_{B3}\nonumber \\
&+&\frac{i}{2}P_L(\frac{2}{\hbar P_L})(W_{1S}-W_{2S}).\hat x^T
e^{{\displaystyle \alpha_+\gamma_+}}h_{++}\psi_{B1}\ =\ 0\ .
\nonumber \\
& &
\end{eqnarray}
\par
These equations allow the elimination of the three components $\psi_{B1}$,
$\psi_{B2}$ and $\psi_{B4}$ in terms of $\psi_{B3}$, which is a surviving
component in the nonrelativistic limit. Defining
\begin{eqnarray} 
\label{5e2}
e^{{\displaystyle 2h_{+-,-+}}} &=& 1-\frac{(m_1-m_2)^2}{P^2}
e^{{\displaystyle -(\alpha_-\gamma_+ + \alpha_+\gamma_-)}}\ \frac
{(f_{+-}-g_{+-})}{(f_{-+}-g_{-+})}\ ,\\
\label{5e3}
e^{{\displaystyle -2u}} &=& e^{{\displaystyle -\alpha_+\gamma_-}}\ 
(f_{+-}-g_{+-})e^{{\displaystyle -2h_{+-,-+}}}\ ,
\end{eqnarray}
and making the wave function transformation
\begin{equation} \label{5e4}
\psi_{B3}\ =\ e^{{\displaystyle u}}\phi_3\ ,
\end{equation}
one obtains the following eigenvalue equation for $\phi_3$, written,
for simplicity, in the c.m. frame:
\begin{eqnarray} \label{5e5}
& & \big \{\ \frac{P^2}{4}\frac{e^{{\displaystyle \alpha_-\gamma_+ +
\alpha_+\gamma_-}}}{(f_{+-}-g_{+-})(f_{-+}-g_{-+})}-\frac{M^2}{4}
\frac{1}{(f_{+-}-g_{+-})^2}-{\bf p}^2\nonumber \\
& &\ \ \ \ \ -\frac{(m_1^2-m_2^2)^2}{4M^2}\frac{1}{(f_{-+}-g_{-+})^2}+
\frac{(m_1^2-m_2^2)^2}{4P^2}\frac{(1+h_{+-}^2)e^{{\displaystyle -(\alpha_-
\gamma_+ + \alpha_+\gamma_-)}}}{(f_{+-}-g_{+-})(f_{-+}-g_{-+})}\nonumber \\
& &\ \ \ \ \ -4\hbar^2{\bf x}^2\ \big [\ u'+\frac{M}{4\hbar r}\frac{h_{+-}}
{(f_{+-}-g_{+-})}-\frac{(m_1-m_2)}{4\hbar r}\frac{h_{-+}}
{(f_{-+}-g_{-+})}\ \big ]^2\nonumber \\
& &\ \ \ \ \ +(6\hbar^2-4{\bf S}^2)\ \big [\ u'+\frac{M}{4\hbar r}\frac
{h_{+-}}{(f_{+-}-g_{+-})}-\frac{(m_1-m_2)}{4\hbar r}\frac{h_{-+}}
{(f_{-+}-g_{-+})}\ \big ]\nonumber \\
& &\ \ \ \ \ +4\hbar^2 {\bf x}^2\ \big [\ u''+\frac{M}{4\hbar}\left(
\frac{h_{+-}}{r(f_{+-}-g_{+-})}\right )'-\frac{(m_1-m_2)}{4\hbar}\left (
\frac{h_{-+}}{r(f_{-+}-g_{-+})}\right )'\ \big ]\ \big \}\ \phi_3\ =\ 0\ .
\nonumber \\
& &
\end{eqnarray}
Here, the prime designates derivation with respect to $r^2$ $(={\bf x}^2)$:
\begin{equation} \label{5e6}
f'\ \equiv \ \frac{\partial f}{\partial r^2}\ ;
\end{equation}
${\bf S}$ is the total spin operator, ${\bf S}={\bf s_1}+{\bf s_2}$,
${\bf S}^2=2\hbar^2s$, $s=0,1$; the other operators and functions are
defined in Eqs. (\ref{3e23})-(\ref{3e24}), (\ref{5e2})-(\ref{5e3}). The
eigenvalues of the matrices $\gamma_{\pm}$ in the sectors with $j=0$ are 
the following:
\begin{eqnarray} \label{5e7}
& & j\ =\ 0\ ,\ \ell \ =\ 0\ ,\ s\ =\ 0\ :\ \ \ \gamma_+\ =\ 3\ ,\ \ \
\gamma_-\ =\ -1\ ;\nonumber \\
& & j\ =\ 0\ ,\ \ell \ =\ 1\ ,\ s\ =\ 1\ :\ \ \ \gamma_+\ =\ -1\ ,\ \ \
\gamma_-\ =\ 3\ .
\end{eqnarray}
The sector with $\ell =0$, $s=0$ corresponds to the pseudoscalar states,
while the sector with $\ell =1$, $s=1$ corresponds to the scalar states.
\par

\newpage

\section{Properties of the eigenvalue equation}
\setcounter{equation}{0}

We study in this section two aspects of the eigenvalue equation
(\ref{5e5}) concerning, first, its short-distance singularities, and, 
second, its nonrelativistic limit.
\par

\subsection{Short-distance singularities}

The question that arises here is whether the presence of the anomalous
magnetic moments has any influence on the short-distance singularities
of the effective potentials present in the eigenvalue equation. It was
already clear from the expressions of the substitutions (\ref{2e11})
that the form factors $b_a(r)$ ($a=1,2$) should be bounded in modulus
by 1 in order not to destabilize at finite distances the bound state
system. A detailed analysis of the eigenvalue equation is however
necessary to reach a more complete understanding of the role of the 
form factors near the origin. We shall limit our study to the case of
Todorov's potential (\ref{2e15}).
\par
For a matter of comparison, we rewrite Eq. (\ref{5e5}) in the case when
the anomalous magnetic moments are absent. Here, we have $B_1=B_2=C=0$,
$A=A_0$, $\alpha_+=\alpha_-=2A_0$, $\beta_+=\beta_-=0$. Denoting
$h\equiv h_{+-,-+}$ [Eq. (\ref{5e2})] in this case, Eq. (\ref{5e5})
becomes \cite{ms1}:
\begin{eqnarray} \label{6e1}
& & \big \{\ \frac{P^2}{4}e^{{\displaystyle 8A_0}}-\frac{M^2}{4}
e^{{\displaystyle 4A_0}}-\frac{(m_1^2-m_2^2)^2}{4M^2}e^{{\displaystyle
4A_0}}+\frac{(m_1^2-m_2^2)^2}{4P^2}\nonumber \\
& & \ \ \ \ \ -{\bf p}^2-4\hbar^2{\bf x}^2h^{\prime 2}+6\hbar^2h'
+4\hbar^2{\bf x}^2h''\nonumber \\
& & \ \ \ \ \ -4{\bf S}^2\ \big [\ (2A_0'+h')(1+4{\bf x}^2A_0')
-(A_0'+2{\bf x}^2A_0'')\ \big ]\ \big \}\ \phi_3\ =\ 0\ .\nonumber \\
& &
\end{eqnarray}
\par
It is sufficient to study the short-distance singularity problem in the 
equal-mass case ($m_1=m_2$, $h=1$) and in the ground state sector 
($\ell =0, s=0$). The dominant singularity comes from the term
$\frac{P^2}{4}e^{8A_0}$, which, according to the expression (\ref{2e15})
of $A_0$, yields the attractive potential $\alpha^2/r^2$. This term is 
at the origin of the fall-to-the-center phenomenon with a critical value
of $\alpha$ equal to $\frac{1}{2}$ \cite{bcs}.
\par
The above analysis can be repeated with Eq. (\ref{5e5}). In the equal-mass
case, one has $b_1=b_2=b$ [Eqs. (\ref{2e11})] and the expressions of the
various potentials [Eqs. (\ref{3e23})-(\ref{3e24}), (\ref{4e2}),
(\ref{4e7})-(\ref{4e9})] become:
\newpage
\begin{eqnarray} \label{6e2}
\alpha_+ &=& \frac{1}{4}\ln\big [\big (1+4v(1-b)^2\big )\big (1+
4v(1+b)^2\big )\big ]\ ,\nonumber \\
\alpha_- &=& \frac{1}{2}\ln\big (1+4v(1-b^2)\big )\ ,\nonumber \\
\beta_+ &=& \frac{1}{4}\ln\big [ \frac{1+4v(1-b)^2}{1+4v(1+b)^2}\big ]
\ ,\ \ \ \ \ \beta_-\ =\ 0\ ,\nonumber \\
f_{+-}-g_{+-} &=& \cosh\sqrt{\alpha_+^2+\beta_+^2\gamma_-^2}-
\frac{\alpha_+}{\sqrt{\alpha_+^2+\beta_+^2\gamma_-^2}}\sinh\sqrt{\alpha_+^2
+\beta_+^2\gamma_-^2}\ ,\nonumber \\
f_{-+}-g_{-+} &=& e^{{\displaystyle -\alpha_-}}\ .
\end{eqnarray}
\par
Their behaviors near the origin are:
\begin{eqnarray} \label{6e3}
\alpha_+ &\simeq& \alpha_-\ \simeq\ \frac{1}{2}\ln(\frac{1}{P_L r})\ ,
\nonumber \\
\beta_+ &\simeq& \frac{1}{4}\ln\left(\frac{1-b}{1+b}\right )^2\ ,
\ \ \ \ 0\leq b(0)<1\ ,\nonumber \\
f_{+-}-g_{+-} &\simeq& e^{{\displaystyle -\alpha_+}}\big (1-\frac
{\beta_+^2\gamma_-^2}{2\alpha_+^2}\big )+\frac{\beta_+^2\gamma_-^2}
{4\alpha_+^2}e^{{\displaystyle \alpha_+}} \nonumber \\
&\simeq& (P_L r)^{1/2}+\beta_+^2\gamma_-^2\big (\ln
(\frac{1}{P_L r})\big )^{-2}(P_L r)^{-1/2}\ ,\nonumber \\
f_{-+}-g_{-+} &\simeq& (P_L r)^{1/2}\ .
\end{eqnarray}
\par
The behavior of ($f_{+-}-g_{+-}$) near the origin crucially depends on 
that of $b(r)$. If $b(0)\neq 0$, then $\beta_+(0)\neq 0$ and hence 
($f_{+-}-g_{+-}$) essentially behaves as $r^{-1/2}$. The first term in
Eq. (\ref{5e5}) has therefore a behavior of the type $r^{-1}$, contrary
to the behavior of the type $r^{-2}$ obtained in the absence of anomalous
magnetic moments. Therefore, a non-vanishing of the form factors $b(r)$
at the origin drastically changes the singularity of the effective 
potential at the origin. Also in this case, for $s=0$, the function $u$
[Eq. (\ref{5e3}) behaves as $-\alpha_+$ and the combination 
$-4\hbar^2{\bf x}^2u^{\prime 2}+6\hbar^2u'+4\hbar^2{\bf x}^2u''$ of Eq. 
(\ref{5e5}) has a behavior close to $\hbar^2/(4r^2)$, which was 
absent in the initial case. This
singularity is independent of the value of the coupling constant $\alpha$
and is located at the critical point. This would mean that the system, 
even for small values of $\alpha$, would face strong attractive 
singularities, which are not observed experimentally.
\par
The above study suggests that the form factors $b(r)$ should vanish at 
the origin, in order not to drastically modify the situation found in the
absence of anomalous magnetic moments. A smooth contribution of the
anomalous magnetic moments would require that the second term in the
right-hand side of the equation of $(f_{+-}-g_{+-})$,
Eq. (\ref{6e3}), be nondominant in front of the first.
This implies that $\beta_+$, and hence $b$, vanish at least as rapidly as
$r^{1/2}$ at the origin. (Also, in this case, the function $u$ vanishes
at the origin.)
\par
A parametrization of $b_a(r)$, corresponding to a vanishing at the origin
as $r$, is obtained with the following choice of the functions $c_a(r)$
of Eqs. (\ref{2e12}):
\begin{equation} \label{6e4}
c_a(r)\ =\ d_a+f_a\frac{\hbar \alpha}{2\pi m_a r}\ \ \ \ \ (a=1,2)\ ,
\end{equation}
with $d_a$ and $f_a$ constants, $d_a>1$, $f_a>0$.
\par

\subsection{Nonrelativistic limit}

When the magnetic moment form factors $b(r)$ are smooth functions, then 
for values of the coupling constant $\alpha$ of the order of $1/2$ (the
critical value), their effects can still be estimated perturbatively,
their order of magnitude being fixed by $\alpha/\pi$. An even cruder
estimate is obtained by the nonrelativistic limit, which allows us 
to have easily an idea of the signs of the energy shifts.
We shall assume that $\alpha$ is sufficiently small to also justify a
perturbative-nonrelativistic treatment of the Coulomb potential $v$
[Eq. (\ref{4e2})] appearing in expressions concerning the anomalous 
magnetic moments.
\par
We treat the form factors $b_a(r)$ to first order.Among the effective
potentials (\ref{3e23})-(\ref{3e24}) and (\ref{4e8})-(\ref{4e9}), only
$\beta_{\pm}$ and the $h$'s are first order quantities in $b_a$. In this
approximation, the latter are given by the perturbation theory result:
\begin{equation} \label{6e5}
b_a(r)\ \simeq \ \frac{\alpha}{2\pi}\frac{1}{2m_a r}\ \ \ \ \ (a=1,2)\ .
\end{equation}
\par
One finds for the effective potentials:
\begin{eqnarray} \label{6e6}
& & h_{+-}\ \simeq\ \beta_+\gamma_-\ ,\ \ \ \ \ h_{-+}\ \simeq\ \beta_-
\gamma_+\ ,\nonumber \\
& & \beta_+\ \simeq\ -2v(b_1+b_2)\ ,\ \ \ \ \ \beta_-\ \simeq\ 
-2v(b_1-b_2)\ .
\end{eqnarray}
\par
Let the first-order perturbation due to the anomalous magnetic moments 
appearing in Eq. (\ref{5e5}) be represented by $-\delta V$. We have:
\begin{eqnarray} \label{6e10}
-\delta V &=& (6\hbar^2-4{\bf S}^2)\ \big [\frac{M}{4\hbar r}h_{+-}-\frac
{(m_1-m_2)}{4\hbar r}h_{-+}\big ]\nonumber \\
& & \ \ \ \ \ +4\hbar^2 {\bf x}^2\ \big [\frac{M}{4\hbar r}h_{+-}-\frac
{(m_1-m_2)}{4\hbar r}h_{-+}\big ]'\nonumber \\
&=&\ \mbox{\boldmath $\nabla$}^2 \int^{{\bf x}^2} dr^2 \big [\frac{M}
{4\hbar r}h_{+-}-\frac{(m_1-m_2)}{4\hbar r}h_{-+}\big ]\nonumber \\
& & -4{\bf S}^2\ \big [\frac{M}{4\hbar r}-\frac{(m_1-m_2)}{4\hbar r}
h_{-+}\big ]\ ,
\end{eqnarray}
where \mbox{\boldmath $\nabla$}$^2$ is the laplacian operator. Using the
expressions of the $h$'s and $b$'s [Eqs. (\ref{6e5})-(\ref{6e6})] we also
have:
\begin{eqnarray} \label{6e7}
\frac{M}{4\hbar r}h_{+-}-\frac{(m_1-m_2)}{4\hbar r}h_{-+} &=& \frac{1}
{4\hbar r}\big [-M2v(b_1+b_2)\gamma_-+(m_1-m_2)2v(b_1-b_2)\gamma_+\big ]
\nonumber \\
&=& -\frac{1}{16\pi \hbar}\frac{1}{m_1m_2M}\frac{\alpha^2}{r^3}\big (
M^2\gamma_-+(m_1-m_2)^2\gamma_+\big )\ .
\end{eqnarray}
Then the corresponding perturbation in the nonrelativistic hamiltonian,
designated by $\delta V_{NR}$, is related to $\delta V$ with the relation
\cite{ms1}:
\begin{equation} \label{6e9}
\delta V\ =\ \frac{2m_1m_2}{M}\delta V_{NR}\ .
\end{equation}
Introducing the total spin quantum number $s$ ($=0,1$), we obtain:
\begin{eqnarray} \label{6e11}
\delta V_{NR} &=&\ \frac{\alpha^2}{4\hbar}\frac{1}{m_1^2m_2^2}\delta^3
({\bf x})[M^2\gamma_-+(m_1-m_2)^2\gamma_+]\nonumber \\
& & -\frac{\alpha^2}{4\pi\hbar}\frac{1}{m_1^2m_2^2}\frac{s}{r^3}
[M^2\gamma_-+(m_1-m_2)^2\gamma_+]\ .
\end{eqnarray}
The first term contributes to the sector with $s=0$, $\ell =0$, for which
$\gamma_+=3$, $\gamma_-=-1$ [Eqs. (\ref{5e7})], while the second one to 
the sector with $s=1$, $\ell =1$, for which $\gamma_+=-1$, $\gamma_-=3$.
The energy shift then becomes:
\begin{eqnarray} \label{6e12}
\delta E &=&\ \frac{\alpha^5}{2\pi}\frac{m_1m_2}{M^3}(m_1^2+m_2^2-
4m_1m_2)\frac{\delta_{\ell 0}\delta_{s0}}{n_{\ell}^3}\nonumber \\
& & - \frac{\alpha^5}{6\pi}\frac{m_1m_2}{M^3}(m_1^2+m_2^2+4m_1m_2)
\frac{\delta_{\ell 1}\delta_{s1}}{n_{\ell}^3}\ ,
\end{eqnarray}
with $n_{\ell}=\ell+n'+1$, $n'=0,1,\ldots$.
\par
While the sign of the energy shift is negative for the sector with
$s=1$, $\ell =1$, it depends on the ratio $m_1/m_2$ for the sector with
$s=0$, $\ell =0$. For the particular case of equal masses, $m_1=m_2$,
the energy shift for the latter sector is negative and equal to the
energy shift of the sector with $s=1$, $\ell =1$.
\par
In the infinite mass limit, $m_2\rightarrow \infty$, the problem reduces
to that of a spin-$\frac{1}{2}$ particle with anomalous magnetic moment
placed in an external static Coulomb field. In this case, the sector
with $s=0$, $\ell =0$ tends to the new sector with $j=\frac{1}{2}$,
$\ell =0$ and the sector with $s=1$, $\ell =1$ to the sector with
$j=\frac{1}{2}$, $\ell =1$. Equations (\ref{6e11}) and (\ref{6e12})
become:
\begin{eqnarray} \label{6e13}
\delta V_{NR} &=& \frac{\alpha^2}{2m_1^2}\delta^3({\bf x})-\frac{\alpha^2}
{2m_1^2}\delta_{\ell 1}\frac{1}{r^3}\ ,\nonumber \\
\delta E &=& m_1\frac{\alpha^5}{2\pi}\frac{\delta_{\ell 0}\delta_{j1/2}}
{n_{\ell}^3}-m_1\frac{\alpha^5}{6\pi}\frac{\delta_{\ell 1}\delta_{j1/2}}
{n_{\ell}^3}\ .
\end{eqnarray}
They agree, as they should, with the corresponding formulas obtained 
directly from the Dirac equation \cite{iz}. (Comparisons of theoretical
predictions involving anomalous magnetic moments with experimental data
can be found in Ref. \cite{lpdr}.)
\par

\newpage

\section{Summary and concluding remarks}
Using a matrix substitution rule, applied to the electric charge and
deduced from the lowest order contribution of the vertex correction in 
QED, we introduced in local form the anomalous magnetic moments at each
vertex of the higher order terms of the constraint theory 
fermion-antifermion interaction potential. Since the latter already has
a local form in three-dimensional $x$-space, determined from summation
of infra-red leading terms of multiphoton exchange diagrams, the new
potential that arises also has a similar locality property and is
calculated by a resummation of the corresponding series after the
incorporation of the anomalous magnetic moments into the vertices.
\par
Focusing our attention to the sectors of pseudoscalar and scalar states
($j=0$), the corresponding wave equations were reduced to a single 
eigenvalue equation. The requirement that the short-distance singularities
of the effective potential should not be drastically enhanced by the 
presence of the anomalous magnetic moments imposed on the accompanying
form factors the condition of a sufficiently rapid vanishing at the 
origin (faster than $r^{1/2}$). It is expected that when this condition is 
realized, then the incorporation of the anomalous magnetic moments, even
in the strong coupling regime, should not introduce qualitative or
destabilizing changes in the properties of the fermion-antifermion bound
state system.
\par

\end{document}